 \documentclass[twocolumn]{aastex631}
\shorttitle{X-ray Flashes on Helium Novae}
\shortauthors{Kato et al.}

\begin{document}

\title{X-Ray Flashes on Helium Novae}

%% Use \author, \affil, and the \and command to format
%% author and affiliation information.
%% Note that \email has replaced the old \authoremail command
%% from AASTeX v4.0. You can use \email to mark an email address
%% anywhere in the paper, not just in the front matter.
%% As in the title, you can use \\ to force line breaks.

\author[0000-0002-8522-8033]{Mariko Kato}
\affil{Department of Astronomy, Keio University, 
Hiyoshi, Kouhoku-ku, Yokohama 223-8521, Japan} 
\email{mariko.kato@hc.st.keio.ac.jp}

%\author{Hideyuki Saio}
%\affil{Astronomical Institute, Graduate School of Science,
%    Tohoku University, Sendai 980-8578, Japan}
% \email{saio@astr.tohoku.ac.jp}

%\and

\author[0000-0002-0884-7404]{Izumi Hachisu}
\affil{Department of Earth Science and Astronomy, 
College of Arts and Sciences, The University of Tokyo,
3-8-1 Komaba, Meguro-ku, Tokyo 153-8902, Japan} 
%%%%%%\email{hachisu@ea.c.u-tokyo.ac.jp}
%%\email{izumi.hachisu@outlook.jp}

%% Notice that each of these authors has alternate affiliations, which
%% are identified by the \altaffilmark after each name.  Specify alternate
%% affiliation information with \altaffiltext, with one command per each
%% affiliation.

%% Mark off your abstract in the ``abstract'' environment. In the manuscript
%% style, abstract will output a Received/Accepted line after the
%% title and affiliation information. No date will appear since the author
%% does not have this information. The dates will be filled in by the
%% editorial office after submission.

\begin{abstract}
A helium nova occurs on a white dwarf (WD) accreting hydrogen-deficient
matter from a helium star companion. 
When the mass of a helium envelope on the WD reaches a critical value,
unstable helium burning ignites to trigger a nova outburst. 
A bright soft X-ray phase appears in an early outbursting phase of a
helium nova before it optically rises toward maximum.  Such an X-ray
bright phase is called the X-ray flash. 
We present theoretical light curves of X-ray flashes for 1.0, 1.2, and
1.35 $M_\sun$ helium novae with mass accretion rates 
of $(1.6-7.5) \times 10^{-7}~M_\sun$ yr$^{-1}$. 
Long durations of the X-ray flashes (100 days to 10 years) and 
high X-ray luminosities ($\sim 10^{38}$ erg s$^{-1}$)
indicate that X-ray flashes are detectable as a new type of X-ray 
transient or persistent X-ray sources.
An X-ray flash is a precursor of optical brightening, so that the
detection of X-ray flashes on helium novae enables us to plan arranged
observation for optical premaximum phases that have been one of
the frontiers of nova study.  We found a candidate object of helium-burning
X-ray flash from literature on extragalactic X-ray surveys. 
This X-ray transient source is consistent with our X-ray flash model
of a $1.35 ~M_\odot$ WD.
\end{abstract}

%% Keywords should appear after the \end{abstract} command. The uncommented
%% example has been keyed in ApJ style. See the instructions to authors
%% for the journal to which you are submitting your paper to determine
%% what keyword punctuation is appropriate.

\keywords{novae, cataclysmic variables --- stars: winds --- stars: X-ray --- white dwarfs}

%% From the front matter, we move on to the body of the paper.
%% In the first two sections, notice the use of the natbib \citep
%% and \citet commands to identify citations.  The citations are
%% tied to the reference list via symbolic KEYs. The KEY corresponds
%% to the KEY in the \bibitem in the reference list below. We have
%% chosen the first three characters of the first author's name plus
%% the last two numeral of the year of publication as our KEY for
%% each reference.

\section{Introduction}
\label{introduction}
A helium nova is a rare phenomenon triggered by unstable helium nuclear 
burning on a white dwarf (WD) \citep{taa80,ibe94,cas98,cui18,kat18hvf}.
The first identified helium nova is the Galactic nova V445 Pup 
\citep{ash03, kat03, iij08}. 
Its outburst was discovered on UT 2000 December 30 \citep{kan00}.

The helium nova as an optical phenomenon was theoretically predicted
by \citet{kat89}: they pointed out two cases of helium accretion; 
(1) the WD accretes helium from a helium star companion, and 
(2) the WD is in a steady hydrogen burning, in which the accreted  
hydrogen is burned into helium at the same rate as the accretion.
The thrid case occurs on a mass increasing WD.  
(3) A helium shell flash eventually occurs after a number of successive 
hydrogen shell flashes and a certain amount of helium accumulates
underneath the hydrogen burning zone \citep[][]{cas98,ida13, kat17shb}.  

Long term evolutions of helium accreting WDs are also interested in binary
evolution models toward a Type Ia supernova \citep{abl22, wan09mc, wan17,
won19, kem21, koo23, mag23}.

%Fig.1
%\placefigure{f1.eps}

\begin{figure*}
\epsscale{1.1}
%%\rotate
%\plotone{hr.m12.ps}
\plotone{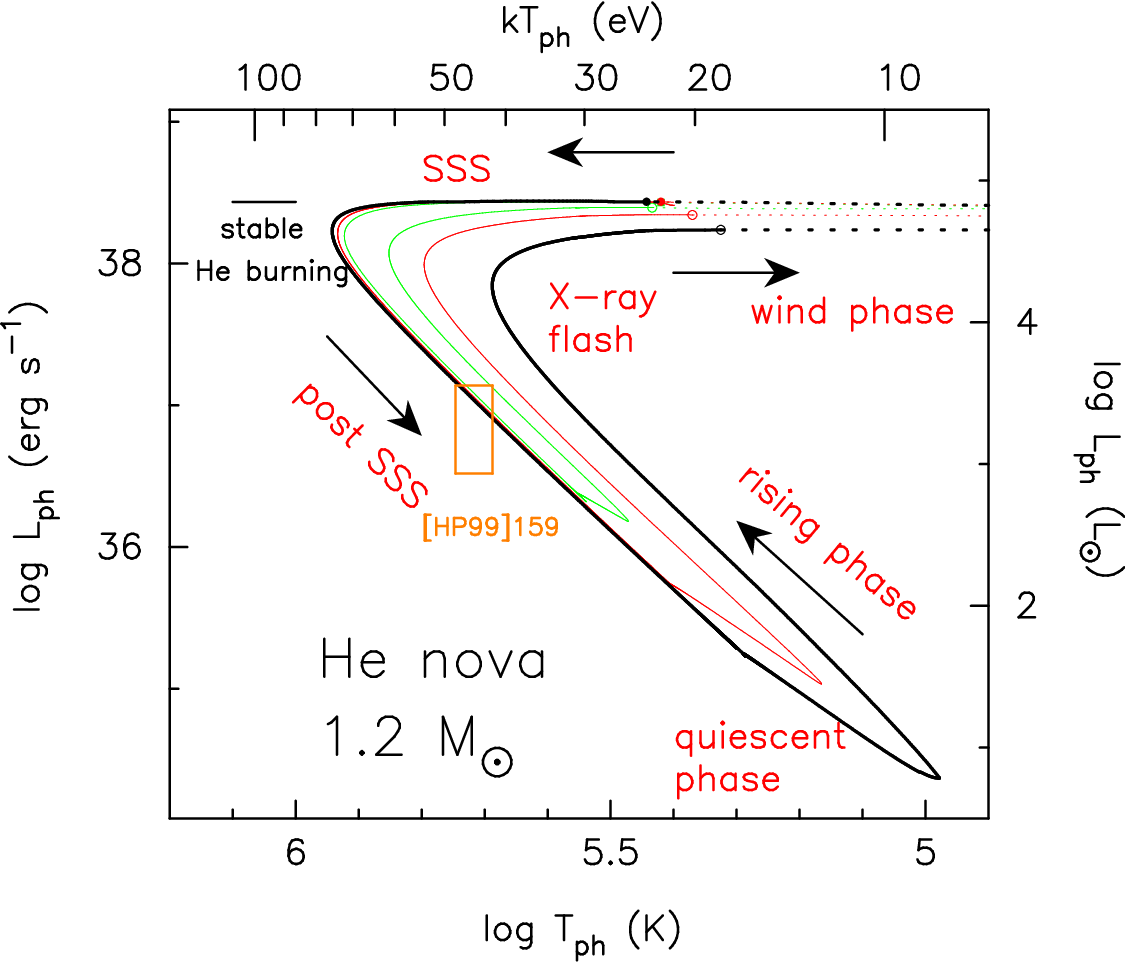}
\caption{
The HR diagram for 1.2 $M_\sun$ helium nova models with 
the mass accretion rate of $\dot{M}_{\rm acc}= 1.6 
\times 10^{-7}$ $M_\sun$ yr$^{-1}$ (black line), 
$3\times 10^{-7}$ $M_\sun$ yr$^{-1}$ (thin red line), and 
$6\times 10^{-7}$ $M_\sun$ yr$^{-1}$ (thin green line). 
The original data are taken from \citet{kat18hvf}. 
The photospheric temperature is denoted by $T$ (K) or $k T$ (eV), 
where $T$ is the temperature in units of Kelvin and 
$k$ is the Boltzmann constant.
Each arrow indicates the direction of evolution.  
Optically thick winds start at the small open circles and ends 
at the filled circles on each track. 
The dotted lines correspond to their wind phases. 
The orange square shows the range of temperature and brightness
for the LMC supersoft X-ray source [HP99]159 estimated by \citet{gre23}. 
The short horizontal bar indicates the luminosity of a 1.2 $M_\sun$ WD
in stable helium burning for the temperature range of [HP99]159.  
}\label{hr}
 % source:  rnhe.mgsi/hr.wip
\end{figure*}

\subsection{Evolutions of helium novae in the HR diagram}
\label{hr_diagram_evolution}  
Figure \ref{hr} shows one cycle helium shell flashes on 
a 1.2 $M_\sun$ WD with the mass accretion rate of $\dot{M}_{\rm acc}= 
1.6$, 3, and $6 \times 10^{-7}$ $M_\sun$ yr$^{-1}$, the data of which 
are taken from \citet{kat18hvf}. 
Each model cycle shows a loop in the HR diagram; it evolves anti-clockwise 
from the bottom (quiescent phase), early X-ray bright phase 
(X-ray flash), wind phase (optically bright phase), supersoft X-ray source
(SSS) phase, and it finally becomes dark (post SSS phase). 

When the envelope expands until the epoch denoted by a small open circle,
optically thick winds begin to emerge from the photosphere.  
Because the winds are accelerated deep inside the photosphere, we 
call them the optically thick winds \citep[e.g.,][]{kat94h}. 
The wind phase begins at the small open circle 
and ends at the filled circle on each track.
In the upper leftward evolution of each track, the envelope mass decreases 
with time owing to both nuclear burning and wind mass loss.  After the wind
stops, the envelope mass decreases owing only to nuclear burning. 
In the SSS phase, the WD photosphere emits supersoft X-rays.  After that,
helium nuclear burning extinguishes and the WD becomes dark. 
The three model tracks are located almost on the same line in the SSS phase 
because each helium envelope reaches the same thermal and hydrostatic balance. 
These physical properties are essentially the same as those in hydrogen 
shell flashes \citep{kov98,den13hb,kat22sh,kat24m1213}.

The short horizontal bar in Figure \ref{hr} corresponds to 
the luminosity of a 1.2 $M_\sun$ WD in the steady helium burning 
$\log L_{\rm ph}/L_\sun =4.85$.  
If the WD accretes mass at the same rate as the mass-decreasing rate 
owing to helium burning, the envelope mass is unchanged and
the photospheric temperature is also the same.  Thus, the WD stays at the 
same position in the HR diagram. 
Such WDs are observationally classified as persistent X-ray sources
\citep{kah97}. 
The mass accretion rates for stable helium burning 
have been obtained for various WD masses 
\citep{kaw88sn, ibe89, wan15, bro16, wan17, kat18hvf}.

\subsection{First helium nova: V445 Puppis}
\label{v445_puppis}  

The spectral feature of V445 Pup
resembles those of classical slow novae except very
strong emission lines of carbon and the absence of hydrogen \citep{iij08}. 
The optical light curve declines very slowly during 200 days
followed by dust blackout, and has not recovered yet. 
We have no report for X-ray detections of V445 Pup.\footnote{No record
of V445 Pup in the ROSAT all sky survey source catalog: period
between June 1990 and August 1991. www.mpe.mpg.de/ROSAT/2RXS} 
Probably a supersoft X-ray source (SSS) phase had been obscured
by a thick dust layer.  Thus, we have no information on the date
when the helium shell burning ended.

\citet{kat08v445pup} calculated theoretical light curve models 
for V445 Pup based on the steady-state approximation, 
and concluded that the WD is as massive as $\gtrsim 1.35~M_\sun$. 
Such a massive WD with a helium star companion has been discussed in 
relation to Type Ia supernova progenitors 
\citep[e.g.,][]{ban03, jac19, koo23, mag23}.

\subsection{Second helium nova: LMC [HP99]159}
\label{hp99_159}  

The LMC supersoft X-ray source [HP99]159 is the second candidate
for a helium nova. \citet{gre23} presented detailed observational summary
for [HP99]159. The optical spectrum of [HP99]159 is hydrogen-free,
so they concluded that the companion is a helium star. 
This object has been observed with several X-ray satellites since 
the Einstein satellite in 1979, followed by 
EXOSAT, ROSAT, XMM-Newton, Swift, and SRG/eROSITA.  
\citet{gre23} estimated the luminosity and temperature of 
[HP99]159 as shown in Figure \ref{hr}. 
They interpreted this object as a steady helium-shell burning source, 
with a suggestion of a 1.2 $M_\sun$ WD, 
although the X-ray flux is two orders of magnitude smaller than 
the luminosity of a steady helium-shell burning WD. 

Note that nuclear burning is stable in the upper leftward 
evolution branch (horizontal part) in the HR diagram (Figure \ref{hr}), 
but unstable in the following downward evolution branch (post-SSS phase)
\citep{sie75, sie80, kaw88sn, nom07}.
When the WD evolves into the lower branch, 
nuclear burning extinguishes and then the WD cools downward on the track. 

\citet{gre23} interpreted the X-ray luminosity of [HP99]159 as 
a steady helium-shell burning on a WD and claimed that such a low 
luminosity is possible when the WD is rotating addressing \citet{yoo04}'s
numerical work.  However, Yoon et al. showed that,  
while rotation does enhance stability,
the accretion rates required to explain LMC [HP99]159 as
a steady-state supersoft X-ray source are still too low to remain
stable even in favorable rotation rates.
This paper rather presents counterargument against such a low-luminosity
steady helium-shell burning. 
Thus, it is unlikely that [HP99]159 is a steady helium shell burning object,
at least, from the theoretical point of view. 

\citet{kat23mnras} presented another interpretation for this faint 
X-ray flux as a post-SSS phase of a helium nova outburst. 
Kato et al. calculated post-SSS light curves based on the helium nova
models, and showed that a $\sim 1.2 ~M_\sun$ helium nova has a long 
decay phase consistent with the X-ray observation of [HP99]159 
since 1980.   
Unfortunately, its optical outburst has not been reported. 
In any case, the presence of a binary with a helium donor star like [HP99]159 
suggests that helium accreting WDs are rather abundant. 

\subsection{Search for X-ray flashes on helium novae}
\label{helium_flash_search}  

The X-ray observation of novae have provided useful information for us
to advance quantitative study of nova outbursts. In particular,
the SSS phase has been a useful tool in determining the WD mass 
\citep[e.g., ][]{sal05,hac10k,wol13, kat20}. 
In helium novae, however, the SSS phase may be difficult to observe 
because a thick dust layer forms soon after the optical maximum,
like in V445 Pup. Extremely strong carbon emission lines observed in V445 Pup
suggest that massive carbonaceous dust forms in helium novae
because carbon is a main nuclear product of helium burning. 
On the other hand, an X-ray flash before optical brightening is not
obscured by dust because of unpolluted environment.
Thus, the detection of X-ray flashes is a unique tool for quantitative 
study of helium novae with X-rays.

In this work, we present theoretical light curves 
of X-ray flashes for helium novae with various WD masses and 
mass accretion rates.
This paper is organized as follows.
Section \ref{section_model} briefly describes our numerical method
and presents results for the X-ray flashes of helium novae.
In Section \ref{transient_sources},
we present a candidate object of helium-burning X-ray flashes. 
Concluding remarks follow in Section \ref{conclusions}.

%Fig.2
\begin{figure*}
\epsscale{0.8}
%%\rotate
%\plotone{hr.he.ps}
\plotone{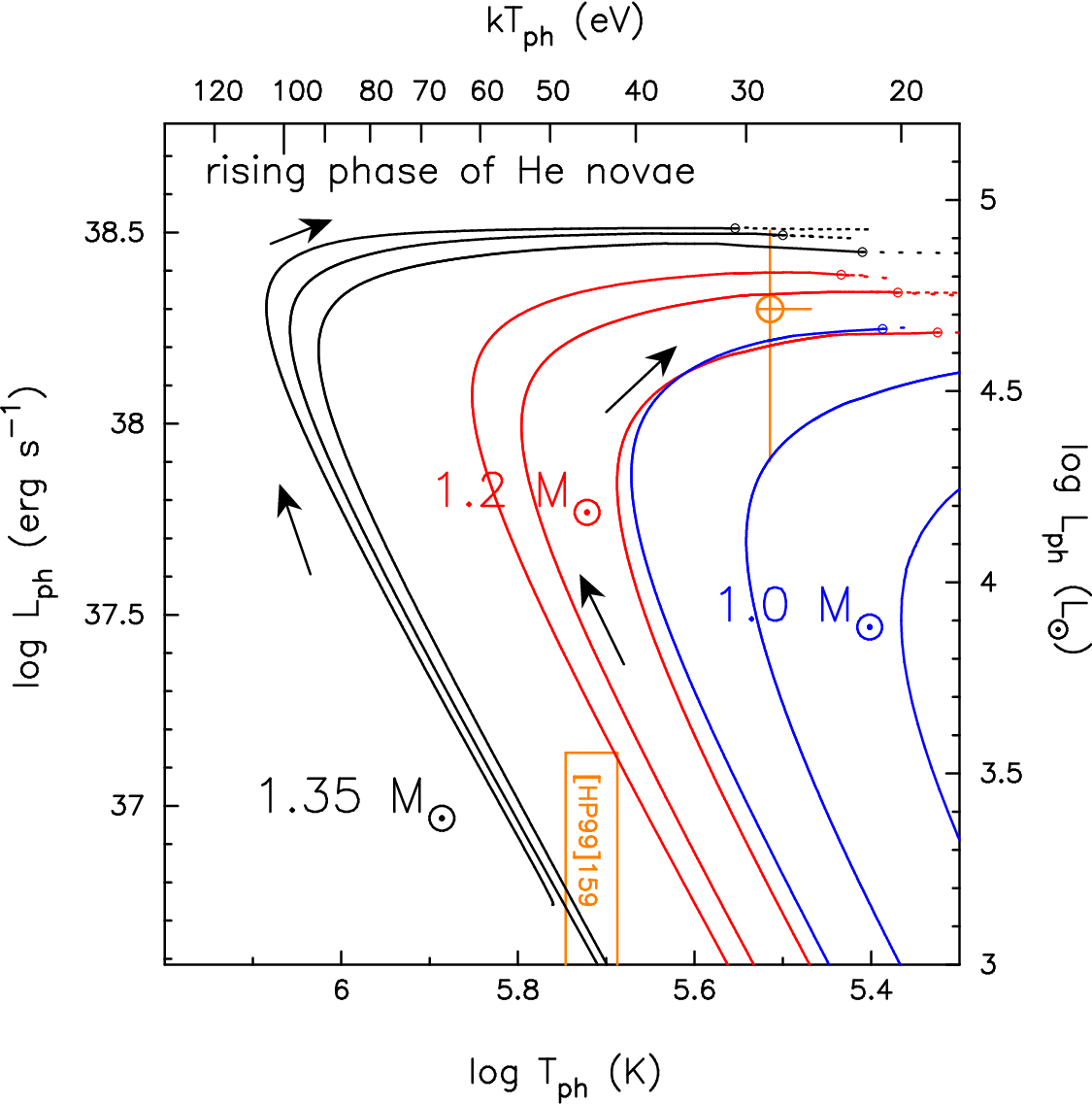}
\caption{
Close up view of rising phases of helium novae in the HR diagram. 
From upper to lower, the three black lines indicate 1.35 $M_\sun$ WD 
models with 
mass accretion rates of $\dot{M}_{\rm acc}= (7.5, 3.0, 1.6)
\times 10^{-7}$ $M_\sun$ yr$^{-1}$,
three red lines 1.2 $M_\sun$ WD models with $(6.0, 3.0, 1.6)\times
10^{-7}$ $M_\sun$ yr$^{-1}$,
and three blue lines 1.0 $M_\sun$ WD models with $(6.0, 3.0, 1.6)\times
10^{-7}$ $M_\sun$ yr$^{-1}$. 
Each arrow indicates the direction of evolution.  
Model parameters are summarized in Table \ref{table_models}.
Optically thick winds start at the small open circles on each track. 
The dotted lines 
corresponds to the wind phase. 
For comparison, we add the position of an X-ray flash of the classical nova
YZ Ret \citep{kon22wa}, which was not a helium nova
but originated from a hydrogen shell flash
(orange circle with error bars). 
The orange square shows the range of temperature and brightness
for the LMC supersoft X-ray source [HP99]159 estimated by \citet{gre23}. 
}\label{hrcompari}
 % source:  nova/yzret/hr.he.wip
\end{figure*}

%Fig.3 
%\placefigure{
\begin{figure*}
\epsscale{0.8}
%\plotone{light.3.ps}
\plotone{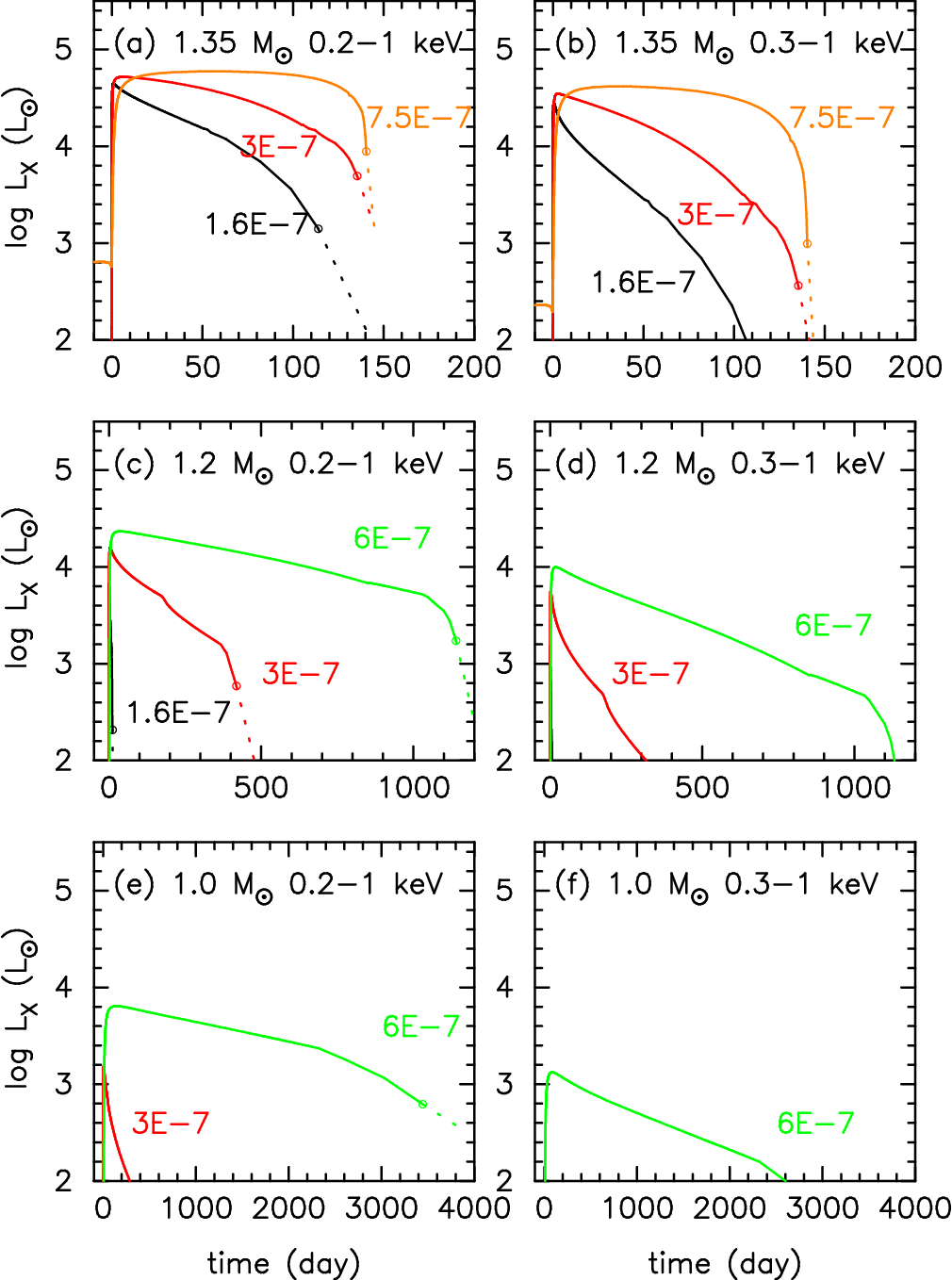}
%%\plottwo{f4a.eps}{f4b.eps}
\caption{
X-ray flash light curves of helium novae.  
The left column panels (a), (c), and (e) show the light curves 
for a $0.2-1.0$ keV energy band, 
corresponding to the SRG/eROSITA instrument. 
The right column panels (b), (d), and (f) show a $0.3-1.0$ keV energy band
for the Swift/XRT.  
In the 1.0 $M_\sun$ WD models, X-ray photons are very soft and
make large difference between the two bands. 
To show that X-ray photons with $E_X > 1$ keV 
are negligible, we recalculated lines in panel (b)
for the energy range of 0.3 - 10 keV. The lines are not changed
within the line width.
The open circles indicate the starting points of optically thick winds
\citep{kat94h}.
Once the winds start to emerge from the photosphere, we expect
no significant X-ray flux owing to the self-absorption effect
of the wind itself outside the photosphere.
\label{light}}
% source rnhe.mgsi/light.3.wip
\end{figure*}

% Table.1
\begin{deluxetable*}{lllrllrr}
\tabletypesize{\scriptsize}
\label{table_models}
\tablecaption{Model parameters of He shell flashes}
\tablewidth{0pt}
\tablehead{
\colhead{Model}&
\colhead{$M_{\rm WD}$}&
\colhead{$\dot M_{\rm acc}$}&
\colhead{$P_{\rm rec}$}&
\colhead{$M_{\rm acc}$}&
\colhead{$L_{\rm nuc}^{\rm max}$}&
\colhead{$\log L^{\rm peak}_{X>0.2}$}&
\colhead{$\log L^{\rm peak}_{X>0.3}$}\\
\colhead{}& 
\colhead{($M_\sun$)}&
\colhead{($M_\sun$ yr$^{-1}$)}&
\colhead{(yr)}&
\colhead{($M_\sun$)}&
\colhead{($L_\sun$)}&
\colhead{($L_\sun$)}&
\colhead{($L_\sun$)}
}
\startdata
M10.6  & 1.0  & 6.0$\times 10^{-7}$  & 1270 &$7.4\times 10^{-4}$ &$2.5 \times 10^{7}$ & 3.81 &3.12\\
M10.3  & 1.0  & 3.0$\times 10^{-7}$ & 7330 &$2.2\times 10^{-3}$  &$6.2 \times 10^{9}$ & 3.18 &2.15\\
M10.16 & 1.0  & 1.6$\times 10^{-7}$ &$31800$ &$5.1\times 10^{-3}$&$2.6 \times 10^{11}$& 2.00 &0.24\\
M10.1  & 1.0  & 1.0$\times 10^{-7}$  &$74200$ & $7.4\times 10^{-3}$&$ 1.5 \times 10^{12}$ & 0.41 & $-2.20$\\
M12.6  & 1.2  & 6.0$\times 10^{-7}$  & 444  &$ 2.6\times 10^{-4}$ &$ 7.1 \times 10^{7}$ & 4.37 & 4.00\\ 
M12.3  & 1.2  & 3.0$\times 10^{-7}$ & 1990 &$ 5.9\times 10^{-4}$ &$3.3 \times 10^{9}$ & 4.20 &3.75\\ 
M12.16 & 1.2  & 1.6$\times 10^{-7}$ & 9340 &$ 1.5\times 10^{-3}$ &$2.5 \times 10^{11}$  &3.83 &3.18\\
M135.75 &1.35 & 7.5$\times 10^{-7}$ & 34.1 &$ 2.5\times 10^{-5}$ &$5.0 \times 10^{6}$& 4.77 & 4.62\\
M135.3  &1.35 & 3.0$\times 10^{-7}$ & 203  &$ 6.0\times 10^{-5}$ & $5.0 \times 10^8 $& 4.72 & 4.54\\
M135.16& 1.35 & 1.6$\times 10^{-7}$ & 743  &$ 1.2\times 10^{-4}$ &$8.5 \times 10^9$ & 4.65 & 4.45
\enddata
%\tablenotetext{a}{The flash duration is defined by the period during which
%$L_{\rm X} > 10^4 L_\sun$.} 
%\tablenotetext{}{A test model with extremely large mass-loss rate.}
\end{deluxetable*}

\section{Model calculation of X-ray flashes}
\label{section_model}

\subsection{X-ray flash light curves on helium novae}
\label{duration}
 
%A nova is a thermonuclear runaway event on a mass-accreting WD 
%\citep{spa78,sio79,nar80, ibe82, pri86, pri95, cas98, taji15,kat22sh}.
%%\citep[see, e.g., ][for a recent self-consistent calculation]{kat22sh}.
%Among them, hydrogen novae have been well studied in both  the
%observational and theoretical points of view, but much less for 
%helium novae as described in Section \ref{introduction}. 

We have calculated X-ray light curves based on the 
helium nova models obtained by \citet{kat18hvf}. 
These models are calculated with a Henyey type evolution code. 
The occurrence of an optically thick wind is detected by the 
condition BC1 in \citet{kat94h}. 
We assumed helium-rich matter accretion of $X=0.0$, $Y=0.98$, and $Z=0.02$. 

Table \ref{table_models} lists our model parameters. 
From left to right, model name, WD mass, mass accretion rate, 
recurrence period of helium nova outbursts, 
accreted mass, maximum nuclear energy generation rate, 
and the maximum X-ray luminosities at the photosphere for 
$0.2 - 1.0$ keV and $0.3 - 1.0$ keV energy band.  

%$\dot{M}_{\rm acc}= 5\times 10^{-10} ~M_\sun$ yr$^{-1}$

These mass-accretion rates are broadly consistent
with the high mass-transfer rates calculated
for Roche-lobe-filling helium stars \citep{kat08v445pup,bro16}. 

%\subsection{X-ray flash phase in the HR diagram}

%X-ray flash light curves have been calculated for hydrogen flashes
%\citep{hil14, kat16xflash, kat22sh, kat22shapjl, kat22shc}, 
%but have not been presented for helium shell flashes.

Figure \ref{hrcompari} is a close-up view of the 
HR diagram that shows each helium nova model 
in Table \ref{table_models}, but 
only in the X-ray flash phase.
For the same WD mass, e.g., $1.35 ~M_\sun$ WD, 
the track of the highest sampled accretion rate is located
on the farthest upper-left side.  As the accretion rate becomes smaller,
the track moves to the lower-right side. 

%The open circle at the right edge of the solid line 
%in Figures \ref{hr} and \ref{hrcompari} 
%indicates the starting point of optically thick winds. 

Figure \ref{light} shows the X-ray light curves during 
the flash phase corresponding to the model in Figure \ref{hrcompari}. 
The left column panels show the $0.2-1.0$ keV band fluxes, corresponding to 
the SRG/eROSITA instrument \citep[e.g.,][]{kon22wa},
while the right column panels are for the $0.3 - 1.0$ keV band,
corresponding to the Swift/XRT \citep[e.g.,][]{eva09}.  
As the X-rays are very soft, most of which are below 1.0 keV,
the flux hardly changes even if we adopt another upper limit of $> 1$ keV. 
We recalculated light curves in Figure \ref{light}b 
for the energy range of 0.3 - 10 keV, but found no differences between
the two different bands.

In each panel, a WD model with a higher mass-accretion rate 
evolves slowly, and the bright X-ray phase lasts longer. 
In a higher mass-accretion rate, the ignition mass is smaller, 
which results in a smaller pressure at the bottom of the envelope. 
Thus, the peak nuclear luminosity 
$L_{\rm nuc}^{\rm max}$ is smaller, as listed in Table \ref{table_models}. 
A smaller nuclear energy generation rate makes expansion slower 
in the X-ray flash phase where the evolution speed is governed by 
convective energy transport.   
Thus, the photospheric radius expands more slowly, and 
the decrease in the photospheric temperature is slower.
This tendency is the same as that in hydrogen shell flashes (hydrogen novae)
as clearly discussed by \citet{kat24m1213}.

In the 1.35 $M_\sun$ models, the X-ray phases last about 150 days 
for the both bands (Figure \ref{light}(a) and (b)).
For less massive WDs, the evolutions are slower and the bright X-ray
phase becomes much longer.  The difference between the two bands
becomes remarkable, especially in the 1.0 $M_\sun$ WD models 
because the emitted supersoft X-rays are softer for lower photospheric
temperatures, as shown in Figure \ref{hrcompari}.

\subsection{Comparison with X-ray flashes on hydrogen novae} 

The classical nova YZ Reticuli is only the nova whose X-ray flash has
been detected \citep{kon22wa}.  It was observed by the eROSITA instrument
on board Spectrum-Roentgen-Gamma (SRG) when it scanned
the region of YZ Ret \citep{kon22wa}.

We plot the position of YZ Ret in the HR diagram (Figure \ref{hrcompari}).
YZ Ret is a hydrogen nova, but it is located at a similar
region to helium novae in the HR diagram. This clearly demonstrates 
that the X-ray flash is detectable also for a helium nova. 

The X-ray flash of YZ Ret lasted very short of $< 8$ hr \citep{kon22wa}
and preceded the optical maximum by about $2-3$ days. 
Such a brief duration and short preceding time to the optical maximum
suggest a very massive WD of $M_{\rm WD} \gtrsim 1.3 ~M_\sun$
\citep{kat22shapjl, kat22shc}. 
\citet{hac23yzret} estimated the WD mass to be 1.33 $M_\sun$ from their 
light curve analysis by comparing model light curves
with multiwavelength observation. 

%{\color{brown}(delete)
%In helium novae, the X-ray flash lasts much longer ($>$ 150 days) 
%than that of a hydrogen-burning X-ray flash case even in a very massive WD
%like $\sim 1.35 ~M_\sun$. 
%If V445 Pup \citep[$\gtrsim 1.35~M_\sun$: ][]{kat08v445pup} were
%exploded in the recent massive X-ray survey era, 
%the X-ray flash could be detected before the optical brightening. 
%If it is the case, we could plan 
%multiwavelength observation for the pre-optical-maximum phase 
%which is one of the last frontiers of nova study. 
%}

The detection of the X-ray flash in YZ Ret gives a large impact 
to the study of novae. 
(1) The luminosity, temperature, and blackbody spectrum \citep{kon22wa} 
are consistent with theoretical scenario that a nova expands 
in almost hydrostatically until an optically thick wind begins to blow
\citep{hac22shock}.
(2) It is confirmed that no strong shock waves form at this very early stage. 
%(3) the counter evidence against such a hypothesis that nova outburst 
%is powered by shocks \citep{li17mc}. 
 
Theoretically, X-ray flash light curves of helium novae should have
similar properties to those in hydrogen novae except for their
timescales \citep[see, e.g.,][for X-ray flash light curves of 
hydrogen novae]{kat22shapjl}. 
We encourage X-ray flash observations of helium novae.

%of supernovae \citep[e.g.,][]{camp07mb}.

\section{Helium-burning X-ray flashes as a new type of transient source}
\label{transient_sources}

%Fig.4
%\placefigure{
\begin{figure}
\epsscale{1.15}
\plotone{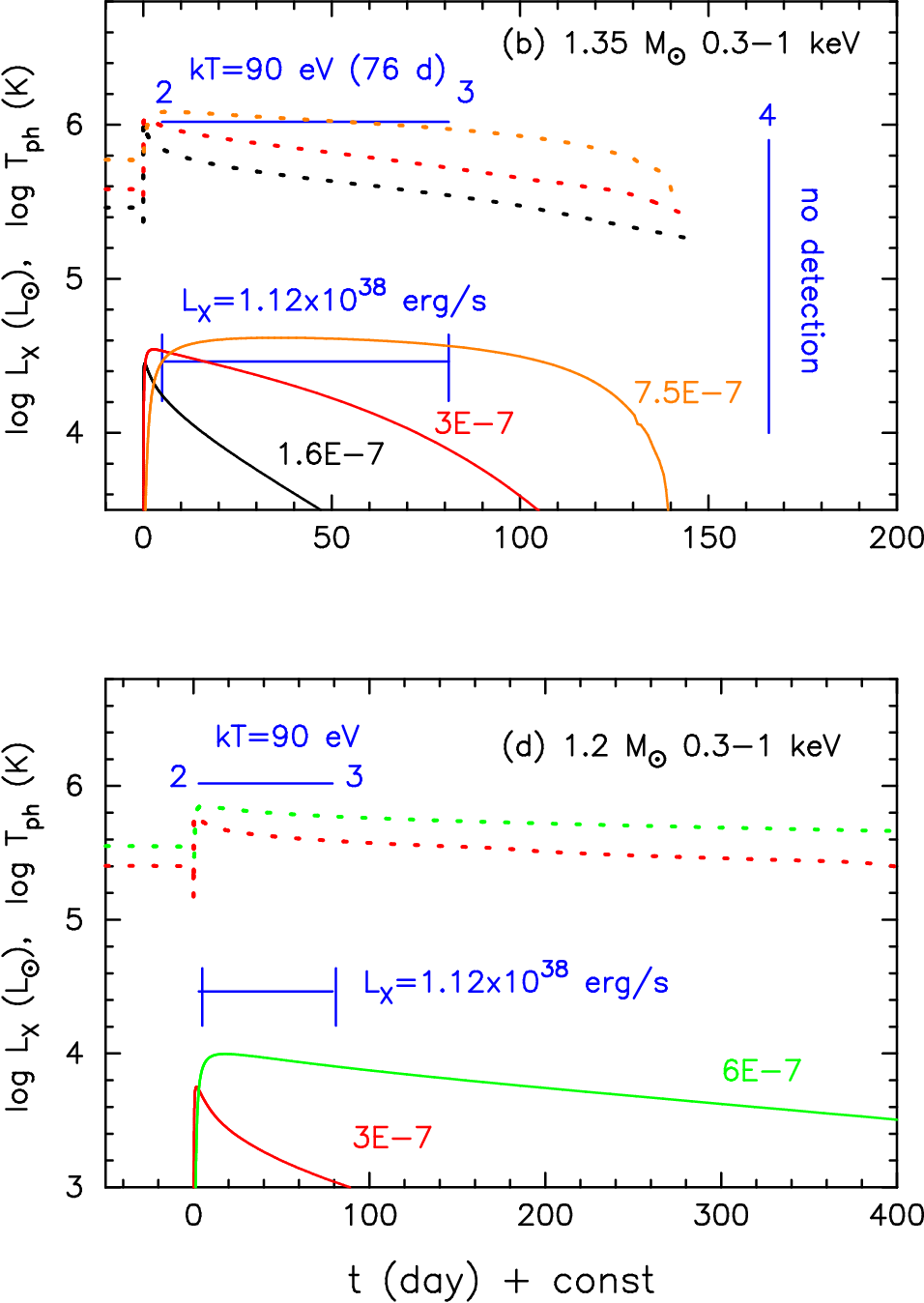}
\caption{
Same as Figure \ref{light}, but only panels (b) and (d) are shown 
with the temporal variation in the photospheric temperature (dotted line). 
The horizontal blue lines show the photospheric blackbody temperature of 
$k T_{\rm ph}= 90$ eV and luminosity $L_{\rm X}= 1.12\times 10^{38}$ erg
s$^{-1}$ with one $\sigma$ range 
estimated for the X-ray source A6 in NGC 3379 \citep{bra12}.
This object is detected twice at epochs 2 and 3
(separated by 76 days), but not at epoch 4. 
\label{lightT}}
% source rnhe.mgsi/light.T.wip
\end{figure}

The X-ray flashes of helium novae could be detected as a new type of
X-ray transient with the X-ray luminosity of $L_{\rm X}\sim 10^{38}$
erg~s$^{-1}$, because of their relatively longer durations (100 days to
10 yr), probably with no optical counterparts.  We searched the literature
of X-ray surveys for candidates of helium-burning X-ray flashes.

\citet{bra08, bra12} reported 18 potential transient X-ray sources 
detected with Chandra in nearby early-type galaxies. 
Among them, the authors classified the X-ray source, A6 
\citep[Source No. 100 in ][]{bra08} in NGC 3379,
to be the SSS phase of a classical nova. 
The X-ray luminosity is $L_{\rm X}= 1.12 \times 10^{38}$ erg~s$^{-1}$ 
(0.3 - 8.3 keV) with a blackbody temperature of 90 eV
with the hydrogen column density of 
$N_{\rm H}=2.8 \times 10^{20}$cm$^{-2}$.
The one $\sigma$ range of $L_{\rm X}$ is 
(0.62 - 1.67)$\times 10^{38}$ erg~s$^{-1}$.  
This source is detected twice with 76 days separation
and did not detected 1805 days before the first detection, 
and 85 days after the second detection.  
No optical counterpart is detected with the limit of 26.4 mag.

It has been discussed that a blackbody fit does not always work well 
in the SSS phase of a nova outburst and results in implausible 
temperature and luminosity \citep[e.g.,][]{kra96, hen11, osb11, pag15}, 
partly because the spectrum is strongly contaminated by absorption and
emission lines.  However, differently to SSS phases, X-ray flashes occur
before massive mass ejection starts.  Therefore, spectra of X-ray flashes
could not be influenced by strong emission lines as observed in the X-ray
flash of YZ Ret. 
%(see discussion in Section \ref{discussion}). 
%As $N_{\rm H}$ is small in A6 as noted above, 
%and thus, we may expect a better fit with a blackbody 
%as in the X-ray flash of YZ Ret \citep{kon22wa}. 
In what follows, we regard that the first selection principle
is the X-ray detection period, 
which is not directly influenced by the ambiguity of blackbody fit.
Then, the second principle is both the luminosity and temperature to avoid
the ambiguity coming from the blackbody fit.  Thus, we selected 
A6 in NGC 3379.

Figure \ref{lightT} shows the X-ray light curves, already shown in Figure
\ref{light}(b) and (d), but we add the temporal variation of the photospheric
temperature (dotted lines).  The blackbody temperature of A6, 
$k T_{\rm ph}= 90$ eV, is
shown in the upper portion of each panel by the horizontal blue line
with a length of 76 days. 
The X-ray luminosity of A6, $L_{\rm X}= 1.12 \times 10^{38}$ erg
s$^{-1}$, is indicated by the horizontal blue line in the lower portion
of the same panel.
We temporarily set epoch 2  at the first positive detection (around day 4), 
correspondingly set epoch 3 at the second detection, and 
epoch 4 at the no detection.  

The $1.35 ~M_\sun$ WD model with the mass accretion rate of
$\dot M_{\rm acc}=7.5 \times 10^{-7} ~M_\sun$ yr$^{-1}$ is consistent with
the observation of A6.  On the other hand, $1.2 ~M_\sun$ WD models show
much lower X-ray luminosities and lower temperatures,
although the flash duration lasts longer. 
In this way, if we have more frequent observation, we can limit 
the range of the WD mass and mass accretion rate.

%Fig.5
%\placefigure{
\begin{figure*}
\epsscale{0.7}
\plotone{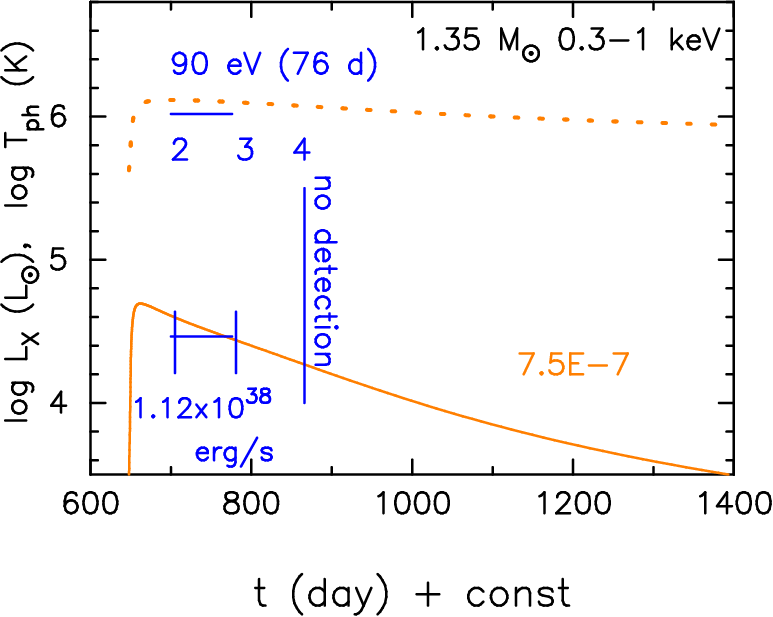}
\caption{
X-ray (0.3 - 1 keV) light curve during the SSS phase of the $1.35~M_\odot$ 
WD with the mass accretion rate of
$\dot M_{\rm acc}=7.5 \times 10^{-7}~M_\sun$ yr$^{-1}$.  
\label{light.T.long}}
% source rnhe.mgsi/light.T.long.wip
\end{figure*}

Next, we examine the possibility of A6 in NGC 3379 to be 
the SSS phase of a helium nova. 
Similar to hydrogen novae \citep{kat22sh,kat24m1213}, 
helium novae also show a much longer SSS phase 
than those of X-ray flashes.  
Figure \ref{light.T.long} shows the SSS light curve 
of the 1.35 $M_\sun$ model with 
$\dot M_{\rm acc}=7.5 \times 10^{-7}~M_\sun$ yr$^{-1}$; 
the SSS phase hardly depends on the mass accretion rate. 
The X-ray light curve shows a gradual decline over several hundred 
days. Epoch 1 (nondetection) is far left outside of this figure, 
which is 1805 days before epoch 2. Early no detection can be 
attributed to the dust blackout, which should have ended before epoch 2. 
The gradual decline of the X-ray light curve is inconsistent 
with nondetection at epoch 4. 

The 1.2 $M_\sun$ models show more gradual decline in the SSS phase
over $> 3000$ days and does not satisfy the observational constrain. 
Thus, it is unlikely that A6 is in the SSS phase of a helium nova.

%Fig.6
%\placefigure{
\begin{figure*}
\epsscale{0.95}
\plotone{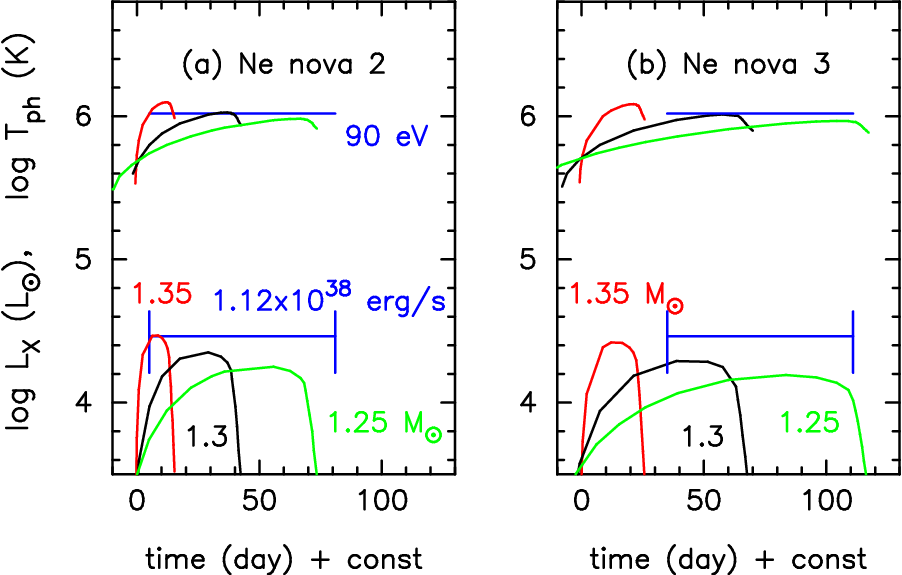}
\caption{
Same as Figure \ref{lightT}, but for the hydrogen-burning SSS phase
(0.3 - 1 keV) of classical novae. 
Each WD mass is attached beside each line; $1.25~M_\odot$ (green line), 
$1.3~M_\odot$ (black line), and $1.35~M_\odot$ (red line). 
The chemical composition is assumed to be (a) Ne nova 2, 
and (b) Ne nova 3 (see the main text). These nova solutions are taken from 
\citet{hac16k}.   
\label{lightT.SSS}}
% source paper/24xflash.he/CN.light.T.wip
\end{figure*}

\citet{bra12} classified the X-ray source A6 in NGC 3379 to be 
a classical nova that is in the SSS phase.  We thus compare the SSS
light curves of classical novae with the A6 observation. 
Figure \ref{lightT.SSS} shows the X-ray light curve (0.3 - 1.0 keV)
and temperature evolution of classical nova models of 
1.25, 1.3, and $1.35 ~M_\odot$ WDs with the chemical compositions 
of the envelope, Neon nova 2 ($X, Y, X_{\rm O},X_{\rm Ne}, Z$)=
(0.55, 0.30, 0.10, 0.03, 0.02), and Neon nova 3 
(0.65, 0.27, 0.03, 0.03, 0.02), which are taken from \citet{hac16k}. 
A nova on a more massive WD evolves faster, and the duration of the SSS phase 
is shorter than 76 days for all the models of Ne nova 2 and 
1.35 and 1.3 $M_\odot$ WD models for Ne nova 3.  
Most of the X-ray photons are emitted below 1.0 keV,
then the X-ray light curves hardly change
even if we adopt a higher upper limit of 8.3 keV corresponding 
to  the Chandra observation.
Thus, five of six models are inconsistent with the Chandra observation 
of A6, although the WD temperature is consistent with the 90 eV.  
Only the 1.25 $M_\odot$ WD (Ne nova 3 composition) shows a longer duration. 
Its flux is smaller than the luminosity of
$L_{\rm X}= 1.12 \times 10^{38}$ erg~s$^{-1}$.
We may find a barely consistent model with the lower one $\sigma$ limit  
between 1.25 $M_\odot$ and 1.3 $M_\odot$ of Ne nova 3 or other composition. 
However, to find a fine tuning model is not our aim of the present work. 

\citet{hen11,hen14} presented statistical relation between the blackbody 
temperature $kT$ and the X-ray turnoff time $t_{\rm off}$ of M31
classical novae.  Here, the X-ray turnoff time is the time of the SSS phase
since the optical outburst, including the optical bright phase.   
The value of $t_{\rm off} \gg$ 76 days with 90 eV is outlier 
of this statistical relation.  
This suggest that A6 is unlikely to be the SSS phase of a classical nova.

X-ray flashes in hydrogen novae are as short as several days or less 
\citep{kat16xflash, kat22sh, kat22shapjl} and, therefore,
are excluded as the explanation of A6.  
From these comparisons, the supersoft X-ray transient 
A6 in NGC 3379 is a candidate of an X-ray flash of a helium nova
on a $\sim 1.35 ~M_\odot$ WD, 
although we cannot completely exclude a possibility of the SSS phase of
a classical nova.

%\section{Discussion}
%\label{discussion}

\section{Concluding remarks}
\label{conclusions}

We have presented theoretical light curves of X-ray flashes for helium novae
and found a candidate object of a helium-burning X-ray flash in the
literature on extragalactic X-ray surveys.  The X-ray flashes last much
longer than those of classical novae, i.e., of hydrogen burnings.  Such
a long duration together with a high X-ray luminosity tells us that X-ray
flashes of helium novae are probably detectable. 
Once an X-ray flash is detected in a helium nova,
we can coordinate multiwavelength observations in 
the following optical rising phase toward maximum.
Because a later SSS phase may not be detectable due to a thick dust blackout, 
the X-ray flash is a valuable tool for us to study helium novae with X-rays.

%\begin{itemize}

%\item 

%\end{itemize}

%% If you wish to include an acknowledgments section in your paper,
%% separate it off from the body of the text using the \acknowledgments
%% command.

%% Included in this acknowledgments section are examples of the
%% AASTeX hypertext markup commands. Use \url without the optional [HREF]
%% argument when you want to print the url directly in the text. Otherwise,
%% use either \url or \anchor, with the HREF as the first argument and the
%% text to be printed in the second.

\begin{acknowledgments}
 We are grateful to the anonymous referees for useful comments,
 which improved the manuscript.
\end{acknowledgments}

\clearpage

%% Use the figure environment and \plotone or \plottwo to include 
%% figures and captions in your electronic submission.

%% If you are not including electronic art with your submission, you may
%% mark up your captions using the \figcaption command. See the 
%% User Guide for details.
%%
%% No more than seven \figcaption commands are allowed per page, 
%% so if you have more than seven captions, insert a \clearpage 
%% after every seventh one. 

%% Tables should be submitted one per page, so put a \clearpage before
%% each one.

%% Two options are available to the author for producing tables:  the
%% deluxetable environment provided by the AASTeX package or the LaTeX
%% table environment.  Use of deluxetable is preferred.
%%

%% Three table samples follow, two marked up in the deluxetable environment,
%% one marked up as a LaTeX table.

%% The following command ends your manuscript. LaTeX will ignore any text
%% that appears after it.

\end{document}